\documentclass[%
 aip,
 apl,
amsmath,
amssymb,
reprint,%
]{revtex4-1}

\usepackage{graphicx}
\usepackage{dcolumn}
\usepackage{bm}

\usepackage[utf8]{inputenc}
\usepackage[T1]{fontenc}
\usepackage{mathptmx}
\usepackage{etoolbox}
\usepackage{makecell}
\usepackage{booktabs}

\makeatletter
\def\@email#1#2{%
 \endgroup
 \patchcmd{\titleblock@produce}
  {\frontmatter@RRAPformat}
  {\frontmatter@RRAPformat{\produce@RRAP{*#1\href{mailto:#2}{#2}}}\frontmatter@RRAPformat}
  {}{}
}%
\makeatother
\begin{document}

\preprint{AIP/123-QED}

\title[Sample title]{Metal-bonded perovskite lead hydride with phonon-mediated superconductivity up to 46 K under atmospheric pressure}

\author{Yong He}
\affiliation{State Key Laboratory for Artificial Microstructures and Mesoscopic Physics, School of Physics, Peking University Yangtze Delta Institute of Optoelectronics, Peking University, Beijing 100871, China}

\author{Juan Du}
 \email{dujuan1121@bjut.edu.cn}
\affiliation{Department of Physics and Optoelectronic Engineering Faculty of Science, Beijing University of Technology, Beijing 100124, China}

\author{Shi-ming Liu}
\affiliation{State Key Laboratory for Artificial Microstructures and Mesoscopic Physics, School of Physics, Peking University Yangtze Delta Institute of Optoelectronics, Peking University, Beijing 100871, China}

\author{Chong Tian}
\affiliation{State Key Laboratory for Artificial Microstructures and Mesoscopic Physics, School of Physics, Peking University Yangtze Delta Institute of Optoelectronics, Peking University, Beijing 100871, China}

\author{Wen-hui Guo}
\affiliation{State Key Laboratory for Artificial Microstructures and Mesoscopic Physics, School of Physics, Peking University Yangtze Delta Institute of Optoelectronics, Peking University, Beijing 100871, China}

\author{Min Zhang}
\affiliation{Inner Mongolia Key Laboratory for Physics and Chemistry of Functional Materials, College of Physics and Electronic Information, Inner Mongolia Normal University, Hohhot 010022, China}

\author{Yao-hui Zhu}
\affiliation{Physics Department, Beijing Technology and Business University, Beijing 100048, China}

\author{Hong-xia Zhong}
\affiliation{School of Mathematics and Physics, China University of Geosciences, Wuhan 430074, China}

\author{Xinqiang Wang}
\affiliation{State Key Laboratory for Artificial Microstructures and Mesoscopic Physics, School of Physics, Peking University Yangtze Delta Institute of Optoelectronics, Peking University, Beijing 100871, China}

\author{Jun-jie Shi}
 \email{jjshi@pku.edu.cn}
\affiliation{State Key Laboratory for Artificial Microstructures and Mesoscopic Physics, School of Physics, Peking University Yangtze Delta Institute of Optoelectronics, Peking University, Beijing 100871, China}

\date{\today}

\begin{abstract}
In the search for high-temperature superconductivity in hydrides, a plethora of  multi-hydrogen superconductors have been theoretically predicted, and some have been synthesized experimentally under ultrahigh pressures of several hundred GPa. However, the impracticality of these high-pressure methods has been a persistent issue. In response, we propose a new approach to achieve high-temperature superconductivity under atmospheric pressure by implanting hydrogen into lead to create a stable few-hydrogen metal-bonded perovskite, Pb$_4$H. This approach diverges from the popular design methodology of multi-hydrogen covalent high critical temperature ($T_c$) superconductors under ultrahigh pressure. By solving the anisotropic Migdal-Eliashberg (ME) equations, we demonstrate that perovskite Pb$_4$H is a typical phonon-mediated superconductor with a $T_c$ of 46 K, which is six times higher than that of bulk Pb (7.22 K) and higher than that of MgB$_2$ (39 K). The high $T_c$ can be attributed to the strong electron-phonon coupling (EPC) strength of 2.45, which arises from hydrogen implantation in lead that induces several high-frequency optical phonon modes with a relatively large phonon linewidth resulting from H atom vibration. The metallic-bonding in perovskite Pb$_4$H not only improves the structural stability but also guarantees better ductility than the widely investigated multi-hydrogen, iron-based, and cuprate superconductors. These results suggest that there is potential for the exploration of new high-temperature superconductors under atmospheric pressure and may reignite interest in their experimental synthesis soon.
\end{abstract}

\maketitle

The discovery of superconductivity with $T_c$=4.15 K in mercury~\cite{Onnes1911} marked the beginning of the arduous quest to identify high-temperature superconductors, which is widely recognized as a grand challenge in condensed matter physics. Based on Bardeen, Cooper, and Schrieffer theory~\cite{Bardeen1957}, Ashcroft hypothesized that the realization of molecular hydrogen metallization under extremely high pressure could lead to the highest $T_c$~\cite{Ashcroft1968}, such as predicted $T_c$ of 764 K in atomic metallic hydrogen under 1-1.5 TPa~\cite{McMahon2011}. However, its metallization in experiments is severely hampered by extremely high-pressure conditions~\cite{McMinis2015, Monserrat2018}. To reduce external pressure, Ashcroft further proposed that chemical precompression can potentially lower the pressure required for hydrogen metallization~\cite{Ashcroft2004}. Specifically, the combination of hydrogen with group IV$\rm _a$ elements can induce precompression, which allows for the transition from an insulating to a metallic phase under highly external pressure~\cite{Ashcroft2004}. Although the electronic properties and phase transition of carbon hydrides under high pressure have been studied by multiple research groups~\cite{Somayazulu1996, Sun2006, Sun2007, He2011}, the discovery of superconductivity in this material has remained elusive. The critical temperature of SiH$_4$ predicted to exhibit superconductivity ranges from 16 K to 106 K under various pressures~\cite{Chen2008, Miguel2009, Zhang2015}. Previous studies have also investigated the superconductivity of germanium and tin hydrides under high pressure~\cite{Zhong2012, Durajski2013, Tse2007, Gao2010, Esfahani2016, Hong2022}. Very recently, Dias's group reported a $T_c$ of 294 K in LuH$_{3-{\delta}}$N$_{\epsilon}$ under about 10000 atmospheres~\cite{Nathan2023}, which quickly caught the attention of the physics community, however, subsequent experimental and theoretical investigations have denied the existence of room-temperature superconductivity in Lu-N-H system~\cite{Xue2023, Xing2023, Huo2023, Xie2023, Sun2023}. These controversies emphasize the need for further experimental and theoretical verification.

It is worth noting that the combination of lead and hydrogen, with an atomic mass ratio of Pb:H=207:1, holds promise as potential superconductors due to the inherent superconducting property of lead~\cite{Floris2007}. Hydrogen implantation has been shown to increase the superconducting critical temperature of lead from 7.22 K~\cite{Boorse1950} to 7.80 K~\cite{Ochmann1981}, with an H/Pb ratio reported to reach 0.20-0.30~\cite{Ochmann1983}. Electron tunneling experiments have indicated that impurity-related localized vibrational modes in hydrogen-doped lead are strongly coupled with the electronic system and vanish at 77 K~\cite{Nedrud1981}. Theoretically, Cheng $et$ $al.$ predicted that the metastable PbH$_4$(H$_2$)$_2$ compound could have a maximum critical temperature of 107 K at 230 GPa, while remaining dynamically and thermodynamically stable above 133 GPa~\cite{Cheng2015}. Chen $et$ $al.$ utilized crystal structure predictions and EPC calculations to report two stable lead hydride compounds, PbH$_6$ with $C222\rm _1$ symmetry and PbH$_8$ with $Fddd$ symmetry, under 200 GPa~\cite{Chen2021}. These compounds exhibit $T_c$ values of 93.33-102.94 K at 100 GPa and 161.59-178.04 K at 200 GPa, respectively. Unlike lighter group-IV$\rm _a$ hydrides such as XH$_4$ (X = C, Si, Ge, and Sn), stable plumbane molecules are not formed in lead hydrides. The current research on lead hydrides primarily focuses on the high-pressure regime, and the structural stability and superconducting properties of lead hydrides under ambient pressure have yet to be studied.

In this letter, we present a comprehensive theoretical investigation into the formation of a stable binary metal-bonded perovskite compound Pb$_4$H at ambient pressure, by combining the heaviest element, Pb, in group-IV$\rm _a$ with H atoms. The H atoms occupy the body-centered site of the face-centered cubic (fcc) Pb atom lattice, which is a complete reversal of the previous approach of synthesizing multi-hydrogen high-$T_c$ superconductors under ultrahigh pressure. Our calculations cover the structural stability, electronic properties, phonon spectrum, EPC, and superconductivity of Pb$_4$H, employing the first-principles methods and ME theory. Our findings reveal that Pb$_4$H exhibits remarkable superconductivity and ductility at ambient pressure, offering valuable insight into potential high-$T_c$ superconductors in few-hydrogen hydrides.

\begin{figure}[ht]
\includegraphics[width=8cm]{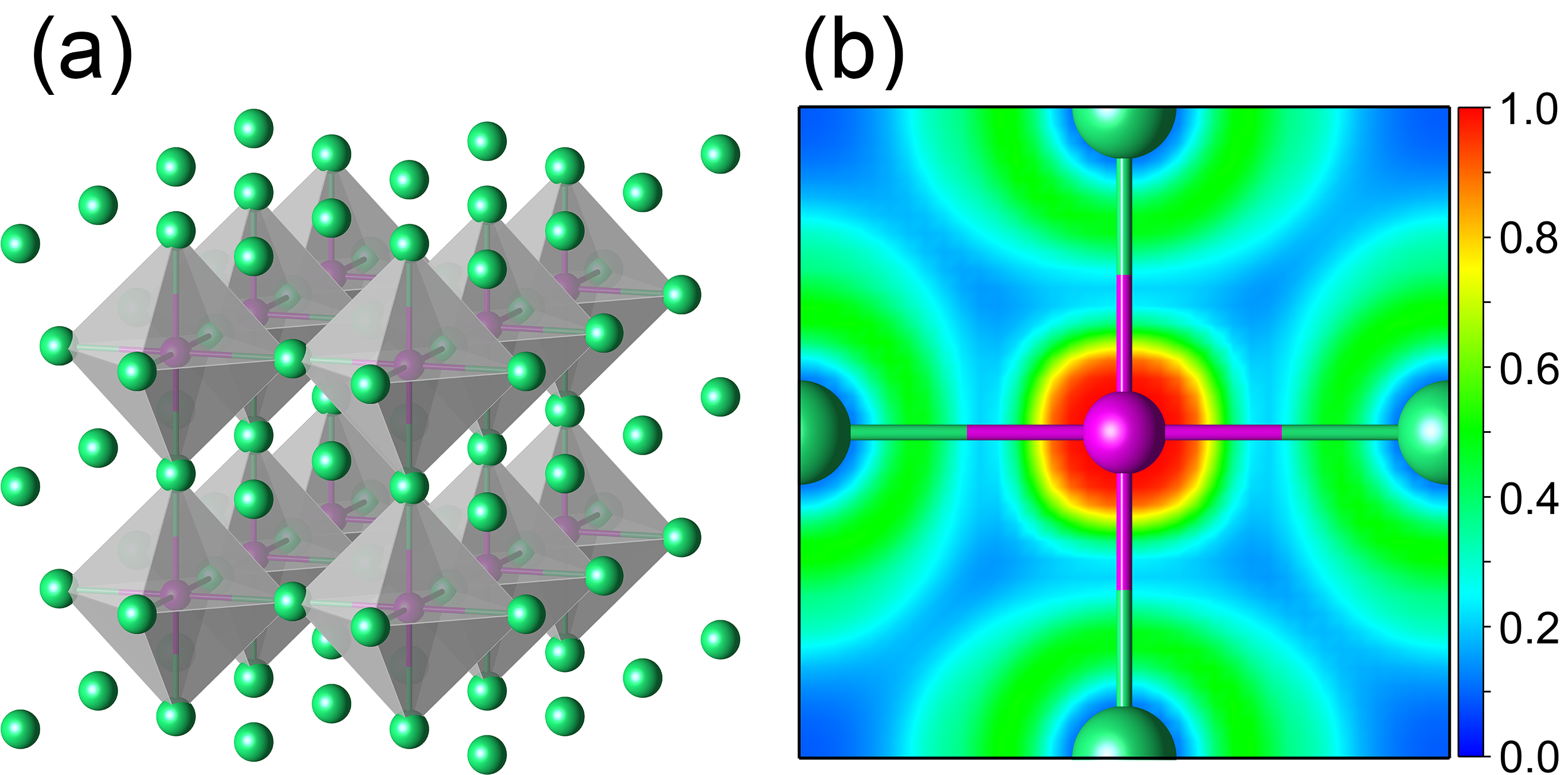}
\caption{\label{Fig_1}Crystal structure and electron localization function (ELF) of Pb$_4$H. (a) The crystal structure of binary metal-bonded perovskite Pb$_4$H, where the green and purple balls represent the Pb and H atoms, respectively. It is clear that the naturally body-centered site, i.e., octahedral interstice (O$\rm _h$), is occupied by the lightest H atom, forming a binary metal-bonded perovskite. (b) Simulated ELF on the Pb-H plane. It can be seen obviously that the bond characteristic between Pb and H atoms is metallic bond, ensuring the favorable ductility of Pb$_4$H.}
\end{figure}

The fcc lattice of lead has an unoccupied body-centered site that can be easily filled by hydrogen due to its small radius, resulting in the formation of binary metal-bonded perovskite Pb$_4$H, as shown in Fig.~\ref{Fig_1} (a), (see Table~S1 and Fig.~S2 for structural details in the supplementary material). In the context of the typical perovskite structure of CaTiO$_3$, this body-centered interstice acts as a normal lattice point, allowing H atom to maintain the structural stability without any external pressure. This design strategy is significantly different from the current approach designing multi-hydrogen high-$T_c$ superconductors under ultrahigh pressure and has the potential for ambient pressure applications. To intuitively reveal the nature of chemical bonds between H and Pb atoms, we conducted the ELF calculations to determine the normalized electron density of perovskite Pb$_4$H. As shown in Fig.~\ref{Fig_1} (b), the background on Pb-H plane is filled with conduction electrons with a normalized electron density of 0.25. Both Pb and H ions are immersed in the electron sea, indicating a typically metallic bond combination, which differs from the covalent or ionic bond found in multi-hydrogen superconductors at ultrahigh pressures. The delocalized electrons are primarily contributed by the 6$s$ and 6$p$ valence electrons of the Pb atom. Based on this analysis, we can infer that the Pb-H metallic bond not only benefits the structural stability of perovskite Pb$_4$H,  but also guarantees its better ductility than multi-hydrogen, iron-based, and cuprate superconductors. More structural stability details are given in the supplementary material.

\begin{figure}[ht]
\includegraphics[width=8cm]{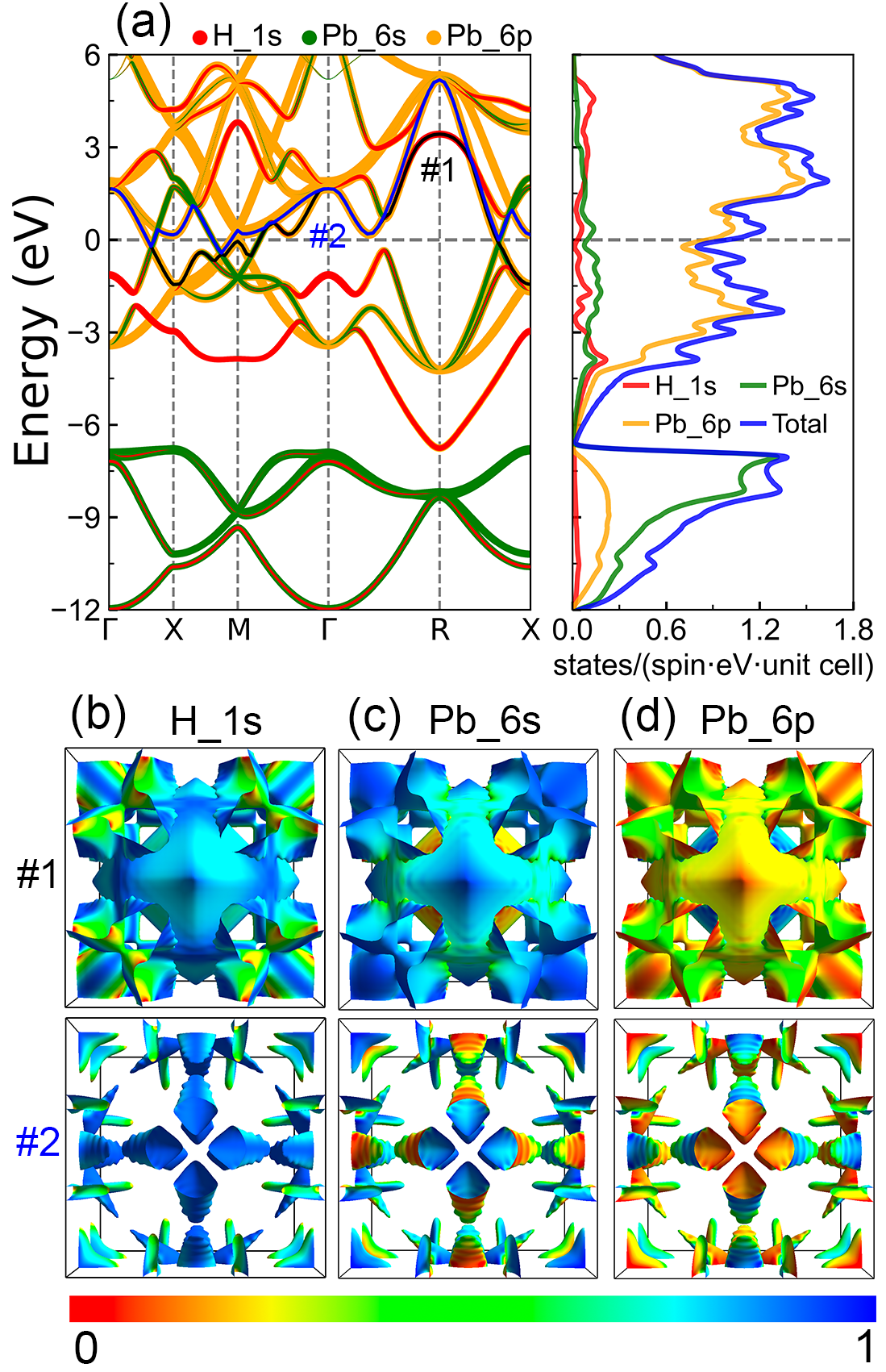}
\caption{\label{Fig_2}Electronic band structure and Fermi surface colored by orbital weight. (a) Projected band structure of Pb$_4$H and corresponding total and projected density of states. The Fermi energy is set to zero. The two bands, labeled as \#1 and \#2, intersect with the Fermi level, indicating a typical metal energy band. The orbital contributions are presented by the size of the colored circles. The distribution of the orbital property on the Fermi surfaces for (b) H 1$s$ orbital, (c) Pb 6$s$ orbital and (d) Pb 6$p$ orbital.}
\end{figure}

We aim to investigate the electronic structures of perovskite Pb$_4$H in more detail. Initially, we calculate the projected band structure, with orbital-resolved contribution, and corresponding total and projected density of states (DOS), depicted in Fig.~\ref{Fig_2} (a). Two bands, labelled as \#1 and \#2, can be observed crossing the Fermi level ($E\rm_F$=0 eV), indicating a clear metallic energy band. The Wannier interpolation bands are presented in Fig.~S1 (a) in the supplementary material. The band \#1, mainly contributed by H 1$s$ and Pb 6$p$ orbitals, crosses $E\rm_F$ along high symmetry point $\Gamma$-X-M-$\Gamma$ and R-X, resulting in a complex Fermi surface (FS) sheet with orbital-resolved contribution, as shown in the top panels of Figs.~\ref{Fig_2} (b)-(d). The band \#2 is primarily derived from the Pb 6$s$ and 6$p$ orbitals, consistent with the FS colored by orbital contribution in the bottom panels of Figs.~\ref{Fig_2} (b)-(d), presenting three intersections with $E\rm_F$. We can observe from the projected DOS that the orbital contribution of Pb 6$p$ is more significant than that of the H 1$s$ state at the $E_F$, which is responsible for the metallic bonding observed in Fig.~\ref{Fig_1} (b). We note that metal Pb also has two bands crossing the $E\rm_F$, which results in a two-band superconductivity~\cite{Floris2007}. Therefore, it is reasonable to infer that Pb$_4$H may also be a two-band superconductor.

\begin{figure}[ht]
\includegraphics[width=8cm]{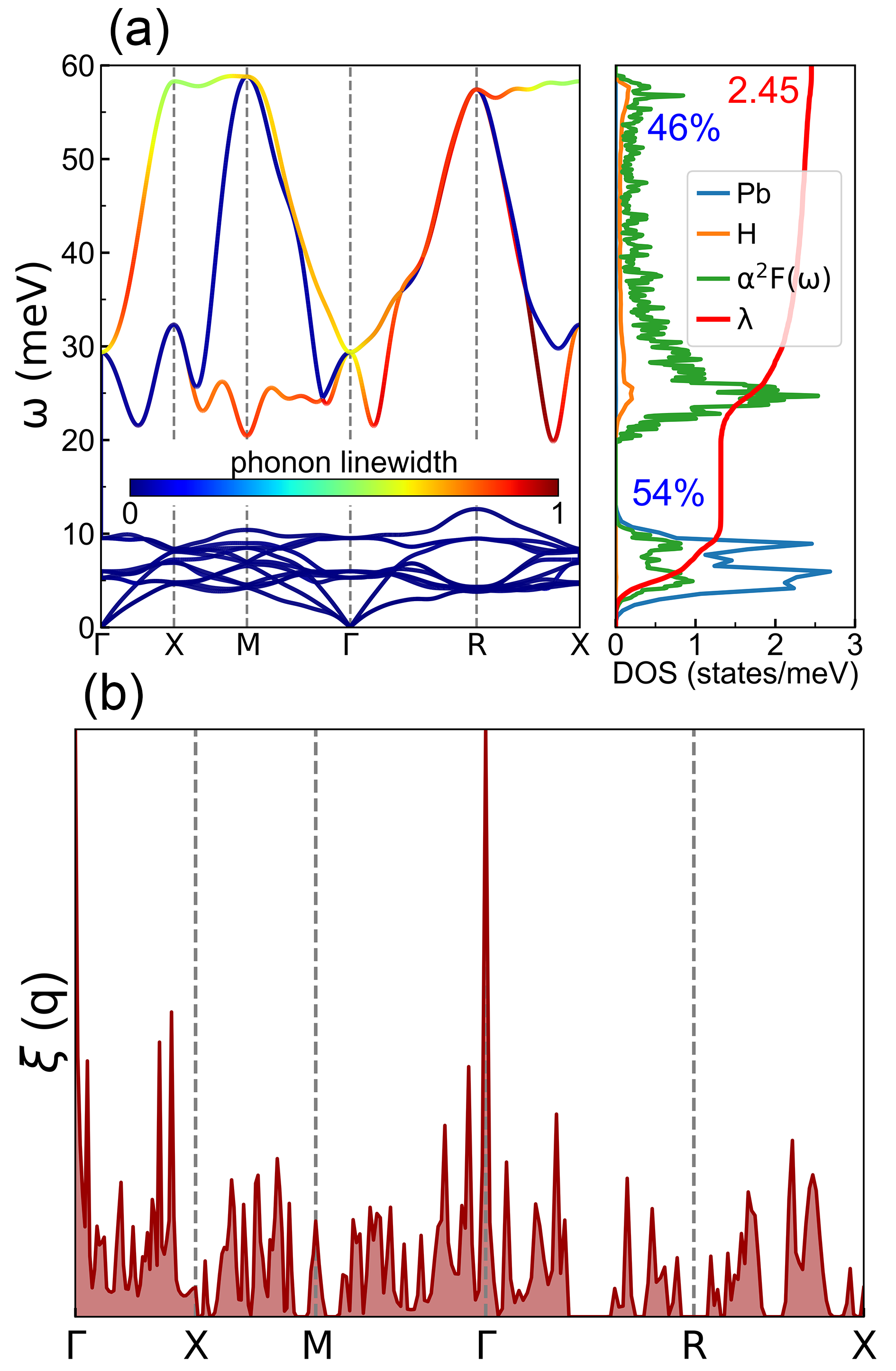}
\caption{\label{Fig_3}Phonon dispersion and FS nesting function. (a) Left: Phonon spectra with the normalized phonon linewidth encoded in color. Right: Projected phonon DOS, Eliashberg spectral function ${\alpha}^2F({\omega})$ and integrated EPC strength $\lambda$. (b) The FS nesting function $\xi$($\mathbf{q}$) along high symmetry path.}
\end{figure}

\begin{figure*}[ht]
\includegraphics[width=16cm]{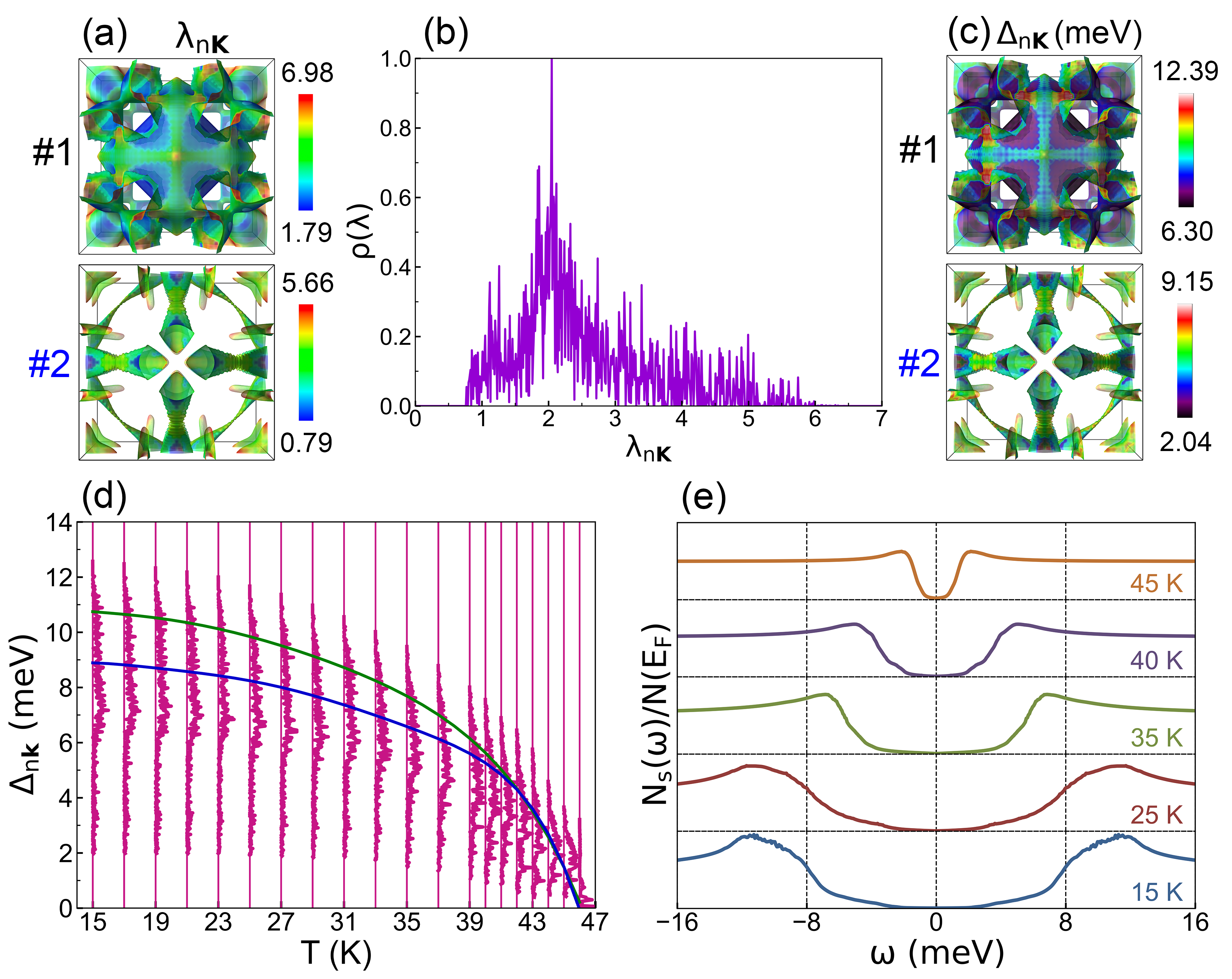}
\caption{\label{Fig_4}EPC strength, superconducting property and quasiparticle DOS of Pb$_4$H. (a) The fully $\mathbf{k}$-resolved EPC constant ${\lambda}_{n\mathbf{k}}$ projected on two FS sheets induced by two crossing bands \#1 and \#2. (b) The normalized distribution of EPC parameter ${\lambda}_{n\mathbf{k}}$. (c) The two FS sheets colored by fully $\mathbf{k}$-resolved superconducting gap ${\Delta}_{n\mathbf{k}}$ at 20 K. (d) Calculated distribution of superconducting gaps ${\Delta}_{n\mathbf{k}}$ versus temperature, where the green and blue solid lines represent the gaps described by the maximum normalized quasiparticle DOS. (e) The normalized quasiparticle DOS from 15 to 45 K.}
\end{figure*}

To investigate the EPC mechanism in perovskite Pb$_4$H, we employed the density functional perturbation theory method~\cite{Baroni2001} to calculate the phonon dispersions. We also used the EPW code~\cite{Giustino2007, Ponce2016, Pizzi2020} for Wannier interpolation on a densely sampled grid of $k$-points and $q$-points. Figure~\ref{Fig_3} (a) shows the phonon spectra with the phonon linewidth indicated by a color bar, as well as the projected phonon DOS, Eliashberg spectrum function ${\alpha}^2F({\omega})$, and the accumulated EPC constant $\lambda$ of Pb$_4$H. The phonon linewidth is determined by the EPC matrix elements and FS nesting function $\xi$($\mathbf{q}$) (see Eq.~(S1)-(S3) in the supplemental material). Specifically, the EPC matrix elements are adopted to reveal the scattering probability amplitude of an electron on the FS sheet through a phonon mode with wave vector $\mathbf{q}$~\cite{Gao2015} and nesting function can be  partially used to describe the EPC strength. As shown in Fig.~\ref{Fig_3} (a), the phonon energy is divided into two parts separated by about 15 meV, owing to the large mass ratio of Pb:H=207:1. The low phonon modes with moderate phonon linewidth contributed by Pb atom vibration are cut off at about 12.62 meV, which is slightly larger than 8.50 meV of bulk Pb~\cite{Noffsinger2010}. On the other hand, the high-frequency branches introduced by H atom vibration exhibit strong electron-phonon interactions with a large phonon linewidth, which plays a significant role in enhancing the critical temperature.

We will now investigate the significant Eliashberg spectral function ${\alpha}^2F({\omega})$ and the accumulated EPC strength $\lambda$ using maximally localized Wannier function. The right panel of Fig.~\ref{Fig_3} (a) shows that the spectral function ${\alpha}^2F({\omega})$ has three major peaks centered at 6, 9, and 25 meV, which are due to the strong electron-phonon interactions represented by a large phonon linewidth. The integrated EPC constant $\lambda$=1.32 contributed by Pb lattice is only slightly smaller than the value of 1.55 obtained from tunneling measurement of bulk Pb~\cite{Rowell1975} and very close to 1.41 obtained by theoretical calculations~\cite{Noffsinger2010}. This indicates that hydrogen implantation in fcc Pb has a negligible effect on electron-phonon interaction dominated by the Pb lattice. The accumulated EPC parameter $\lambda$=1.13, derived from the H atom, accounts for about 46\% of the total $\lambda$, indicating an important positive contribution for improving the EPC strength. The total $\lambda$=2.45 of Pb$_4$H has an enhancement of 58\% compared with the value of 1.55 of bulk Pb~\cite{Rowell1975}, and is about three times larger than the value of 0.75 of MgB$_2$~\cite{Margine2013}, which is due to the large phonon linewidth of the high-frequency phonon branches introduced by the H atom vibration. It is noteworthy that the quite large $\lambda$=2.45 of Pb$_4$H under ambient pressure is similar to the value of 2.29 of LaH$_{10}$ at 250 GPa~\cite{Liu2017} and 2.50 of ThH$_{10}$ at 100 GPa~\cite{Kvashnin2018}. This provides a typical model for investigating strong electron-phonon interaction under ambient pressure.

To gain a deeper understanding of the strong EPC mechanism in Pb$_4$H, we performed additional calculations of the nesting function $\xi$($\mathbf{q}$) along the high-symmetry path, as depicted in Fig.~\ref{Fig_3} (b). Notably, the highest value of $\xi$($\mathbf{q}$) at the $\Gamma$ point lacks a physical interpretation because the entire FS sheet nests into itself. Along the entire high-symmetry direction, numerous sharp peaks are observed, which quantitatively confirm the presence of strong FS nesting at these directions. Interestingly, the soft phonon modes with large phonon linewidth are localized in the nesting regions, indicating that the strong electron-phonon interactions partially originate from FS nesting.

To investigate how the anisotropic EPC strength $\lambda$ in each FS sheet affects the superconductivity of Pb$_4$H, we examine the $\mathbf{k}$-resolved EPC strength ${\lambda}_{n\mathbf{k}}$, which is calculated using the formula $\lambda_{n \mathbf{k}}=\sum_{m \mathbf{k}^{\prime}, \nu} \frac{1}{\omega_{\mathbf{q} v}} \delta\left(\epsilon_{m \mathbf{k}^{\prime}}\right)\left|g_{n \mathbf{k}, m \mathbf{k}^{\prime}}^{\nu}\right|^{2}$~\cite{Margine2013}. Figure~\ref{Fig_4} (a) shows the distribution of ${\lambda}_{n\mathbf{k}}$ on two FS sheets, with values of 1.79-6.98 on FS \#1 and 0.79-5.66 on FS \#2. The normalized distribution of ${\lambda}_{n\mathbf{k}}$ in Fig.~\ref{Fig_4} (b) reveals a wide range of values from 0.77 to 6.59, indicating strong anisotropy within each FS sheet. In particular, the red area on FS \#1 exhibits the largest EPC strength due to the strong coupling of H 1$s$ and Pb 6$p$ electrons with phonons (see top panels of Fig.~\ref{Fig_2} (b) and (d)). Conversely, the EPC strength ${\lambda}_{n\mathbf{k}}$ associated with FS \#2 is primarily derived from Pb 6$s$ and 6$p$ orbitals (see bottom panels of Fig.~\ref{Fig_2} (c) and (d)). The strong anisotropy of EPC strength can enhance the $T_c$~\cite{Floris2007}, making it an important factor in understanding the superconductivity of Pb$_4$H.

After analyzing the EPC strength, we investigated the anisotropic superconducting gaps ${\Delta}_{n\mathbf{k}}$ inside each FS sheet using anisotropic ME equations~\cite{Margine2013} with ${\mu}^*$=0.10. As shown in Fig.~\ref{Fig_4} (c), the superconducting gap ${\Delta}_{n\mathbf{k}}$ at 20 K is decorated on the two FS sheets. The gaps ${\Delta}_{n\mathbf{k}}$ on FS \#1 and \#2 have ranges of 6.30-12.39 and 2.04-9.15 meV, respectively, indicating strong anisotropy of ${\Delta}_{n\mathbf{k}}$ on each FS sheet. This observation is verified by the pronounced vertical energy spread shown in Fig.~\ref{Fig_4} (d). Interestingly, we have observed that the distribution of superconducting gap ${\Delta}_{n\mathbf{k}}$ on FSs is strikingly similar to that of EPC strength ${\lambda}_{n\mathbf{k}}$, indicating that perovskite Pb$_4$H can be regarded as a typical phonon-mediated high-temperature superconductor. As a roughly estimate, the $T_c$ of Pb$_4$H is determined to be 28.16 K by using Allen-Dynes-modified McMillan equations (see Eq.~(S7)-(S11) in the supplementary material).

In the subsequent analysis, we examine the energy distribution of superconducting gaps ${\Delta}_{n\mathbf{k}}$ versus temperature, as presented in Fig.~\ref{Fig_4} (d). The two superconducting gaps derived from the anisotropic nature of the two FS sheets overlap with each other due to the contiguous distribution of ${\lambda}_{n\mathbf{k}}$, where $n$ is the band index of the two bands crossing the $E\rm _F$. This is different from metal Pb where two superconducting gaps are distinctly separated~\cite{Floris2007}. The green and blue solid lines represent the gaps closely related to the superconducting DOS on FS \#1 and \#2, respectively (see Fig.~\ref{Fig_4} (e)). By defining the critical temperature at T=$T_c$ with ${\Delta}_{n\mathbf{k}}$=0, we determine the $T_c$ of Pb$_4$H to be 46 K, which is approximately six times larger than that of lead (7.22 K)~\cite{Boorse1950}, twice as large as Nb$_3$Ge (23 K)~\cite{Testardi1974}, and higher than that of MgB$_2$ (39 K)~\cite{Margine2013}. To assess the significance of the anisotropic effect, we also investigate the temperature dependence of isotropic superconducting gaps ${\Delta}_{\mathbf{k}}$ by solving the isotropic ME equations with ${\mu}^*$=0.10. As illustrated in Fig.~S5 (a) in the supplemental material, the calculated critical temperature is 38 K. Clearly, the presence of multiband and anisotropic gaps results in a 21\% improvement in anisotropic superconductivity relative to isotropic superconductivity. Given the strong electron-phonon interaction featured by large EPC strength $\lambda$=2.45 in Pb$_4$H, we further investigate the $T_c$ by using ${\mu}^*$=0.13. Figure~S5 (b) in the supplemental material shows that the calculated $T_c$ reaches 44 K, indicating that Pb$_4$H is indeed a high-$T_c$ superconductor. 

The quasiparticle DOS (QPDOS) is a critical parameter that can be observed directly using tunnel-conductance measurements. To investigate this parameter in perovskite Pb$_4$H, we calculate the normalized QPDOS in the superconducting state $N_S(\omega)$ at different temperatures using the formula $\frac{N_{S}(\omega)}{N({E\rm_F})}=\operatorname{Re}[\frac{\omega}{\sqrt{\omega^{2}-\Delta^{2}(\omega)}}]$~\cite{Margine2013}, where $N({E\rm_F})$ is the DOS at the Fermi level in the normal state. As shown in Fig.~\ref{Fig_4} (e), the normalized QPDOS displays two peaks, corresponding to the two different energy gap distributions in Pb$_4$H. The superconducting gap decreases with increasing temperature and eventually vanishes in the normal state. Finally, the specific heat and critical magnetic field are investigated, more details are presented in the supplementary material.

In summary, this letter presents a comprehensive investigation of hydrogen-implanted lead and reports the discovery of a new and stable binary metal-bonded perovskite, Pb$_4$H, which exhibits excellent superconductivity and ductility. The investigation utilizes first-principles calculations and Wannier interpolation combined with ME equations. The Pb$_4$H exhibits a $T_c$ of up to 46 K, which is six times higher than that of bulk Pb (7.22 K) and higher than that of MgB$_2$ (39 K). The high-temperature superconductivity of Pb$_4$H can be attributed to an enhancement of EPC strength by 58\% compared with bulk Pb. This improvement arises from the introduction of new high-frequency optical phonon modes resulting from H atom vibration with a relatively large phonon linewidth. Furthermore, the metallic-bonding in perovskite Pb$_4$H guarantees its superior ductility compared with other multi-hydrogen, iron-based, and cuprate superconductors. Importantly, this investigation provides a new design strategy for few-hydrogen metal-bonded hydride high-temperature superconductors with a simple structure under atmospheric pressure, which is in contrast to the well-accepted design method of multi-hydrogen covalent high-$T_c$ superconductors under ultrahigh pressure. These findings are significant in rekindling enthusiasm for predicting new materials with favorable superconducting properties among hydrides at ambient pressure, and they can inspire further experimental realizations in the near future.

See the supplementary material for more information on computational details, structure and stability, superconducting property, specific heat and critical magnetic field.

This work was supported by the Beijing Outstanding Young Scientist Program (BJJWZYJH0120191000103), the National Natural Science Foundation of China (12247143, 12104421, 61734001, 11947218 and 61521004), Zhejiang Provincial Natural Science Foundation of China (LY23A040005), Knowledge Innovation Program of Wuhan-Shuguang Project (2022010801020214), and China Postdoctoral Science Foundation (2022M712965). We used the High Performance Computing Platform of the Center for Life Science of Peking University.

\section*{Author Declarations}
\subsection*{Conflict of Interest}
The authors declare no conflicts of interest.

\subsection*{Author Contributions}
\noindent 
\textbf{Yong He:} Conceptualization (supporting); Data curation (equal); Formal analysis (lead); Visualization (lead); Writing – original draft (lead); Writing – review \& editing (equal). \textbf{Juan Du:} Supervision (supporting); Visualization (supporting); Writing – review \& editing (equal). \textbf{Shi-ming Liu:} Data curation (supporting); Formal analysis (supporting). \textbf{Chong Tian:} Data curation (supporting); Formal analysis (supporting). \textbf{Wen-hui Guo:} Data curation (supporting); Formal analysis (supporting). \textbf{Min Zhang:} Formal analysis (supporting); Visualization (supporting). \textbf{Yao-hui Zhu:} Formal analysis (supporting); Data curation (supporting). \textbf{Hong-xia Zhong:} Formal analysis (supporting); Funding acquisition (equal). \textbf{Xinqiang Wang:} Funding acquisition (equal). \textbf{Jun-jie Shi:} Conceptualization (lead); Supervision (lead); Visualization (supporting); Writing – review \& editing (equal).

\section*{Data Availability}
The data that support the findings of this study are available from the corresponding author upon reasonable request.

\section*{References}

\providecommand{\noopsort}[1]{}\providecommand{\singleletter}[1]{#1}%

\end{document}